\documentclass{iau} 
\usepackage{graphicx}

\title[S318.~NELIOTA: ESA's new NEO lunar impact monitoring project] 
{NELIOTA: ESA's new NEO lunar impact monitoring project with the 1.2m
  telescope at the National Observatory of Athens}

\author[Bonanos et al.]  
       {A.Z. Bonanos$^1$, M. Xilouris$^1$, P. Boumis$^1$, I.
         Bellas-Velidis$^1$, A. Maroussis$^1$, A. Dapergolas$^1$,
         A. Fytsilis$^1$, \\ V. Charmandaris$^1$, K. Tsiganis$^2$,
         K. Tsinganos$^3$}

\affiliation{$^1$IAASARS, National Observatory of Athens, 15236 Penteli,
  Greece, email: {\tt bonanos@noa.gr} \\[\affilskip] $^2$Aristotle
  University of Thessaloniki, 54124 Thessaloniki, Greece
  \\[\affilskip]$^3$University of Athens, Panepistemiopolis, 15784
  Zografos, Athens, Greece}

\pubyear{2015}
\volume{318}  
\jname{Asteroids: New Observations, New Models}
\editors{S. Chesley, A. Morbidelli, R. Jedicke, \& D. Farnocchia, eds.}
\begin{document}

\maketitle

\begin{abstract}
NELIOTA is a new ESA activity launched at the National Observatory of
Athens in February 2015 aiming to determine the distribution and
frequency of small near-earth objects (NEOs) via lunar monitoring. The
project involves upgrading the 1.2m Kryoneri telescope of the National
Observatory of Athens, buliding a two fast-frame camera instrument, and
developing a software system, which will control the telescope and the
cameras, process the images and automatically detect NEO
impacts. NELIOTA will provide a web-based user interface, where the
impact events will be reported and made available to the scientific
community and the general public. The objective of this 3.5 year
activity is to design, develop and implement a highly automated lunar
monitoring system, which will conduct an observing campaign for 2 years
in search of NEO impact flashes on the Moon. The impact events will be
verified, characterised and reported. The 1.2m telescope will be capable
of detecting flashes much fainter than current, small-aperture, lunar
monitoring telescopes. NELIOTA is therefore expected to characterise the
frequency and distribution of NEOs weighing as little as a few grams.
\end{abstract}

\vspace{0.5cm}

Near-Earth Objects (NEOs) are ubiquitous in the space environment. They
are thought to originate from fragments created during asteroid
collisions, asteroids diverted from the asteroid belt through the
gravitational influence of planets, or cometary debris. NEOs have orbits
crossing into the inner Solar System and intersecting the Earth's
trajectory, posing a threat to artificial satellites, spacecraft, and
astronauts. The atmosphere of the Earth offers protection from all but
the largest NEO impacts, which do not completely burn up as they enter
the atmosphere, at speeds of tens of km s$^{-1}$. However, the surface
of the Moon remains susceptible to impacts by small NEOs and can be used
to study their properties.
 
NEO lunar impacts are observed as bright flashes of light. The impacting
meteoroids travel at large speeds (20 to 50 km s$^{-1}$), and thus
contain tremendous kinetic energy that causes the rocks and soil on the
lunar surface to heat up and glow. Ground-based observers detect flashes
lasting from a fraction of a second to several seconds, with light
curves showing a sharp rise and an exponential fading tail. Surveys show
a peak in impacts during meteor showers, as the Earth-Moon system passes
through relatively dense clouds of meteoroids, when crossing the orbits
of comets, however, impacts are detected continuously, without them
necessarily exhibiting a connection to comet debris or a meteor shower.
 
In order to quantify the frequency and characteristics of NEOs, several
campaigns are underway, such as the Lunar Impact Monitoring at NASA's
Marshall Space Flight Center (Suggs et al. 2014), the MIDAS project
(Madiedo et al. 2014, 2015), and the ILIAD Network (Ait Moulay Larbi et
al. 2015). Suggs et al. (2014) reported over 300 impacts down to $R=
10.2$ mag, while surveying an area of 3.8 x 10$^6$ km$^2$ over 7
years. The analysis of 126 flashes that were detected during photometric
conditions, yielded a survey completeness limit of $R= 9$ mag. The
association of certain impacts with meteor streams provided constraints
on the impact speeds and thus their kinetic energy.

Lunar monitoring surveys for NEO impacts typically involve small, 30-50
cm telescopes, tracking at the lunar rate, that are equipped with video
cameras recording at a rate of 30 frames per second (fps). The dark
portion of the lunar surface is monitored during the phases
corresponding to 10-50\% illumination. The aim is to maximize the number
of lunar impacts detected, by maximizing the lunar surface observed,
while avoiding the illuminated surface of the Moon. The goal of such
surveys is to measure the distribution of sizes and masses of objects
impacting the Moon, as well as their flux, and detect a significant
number of impacts from which to obtain statistical results on their
characteristics.
 
\begin{figure}[bh]
\begin{center}
 \includegraphics[width=11cm]{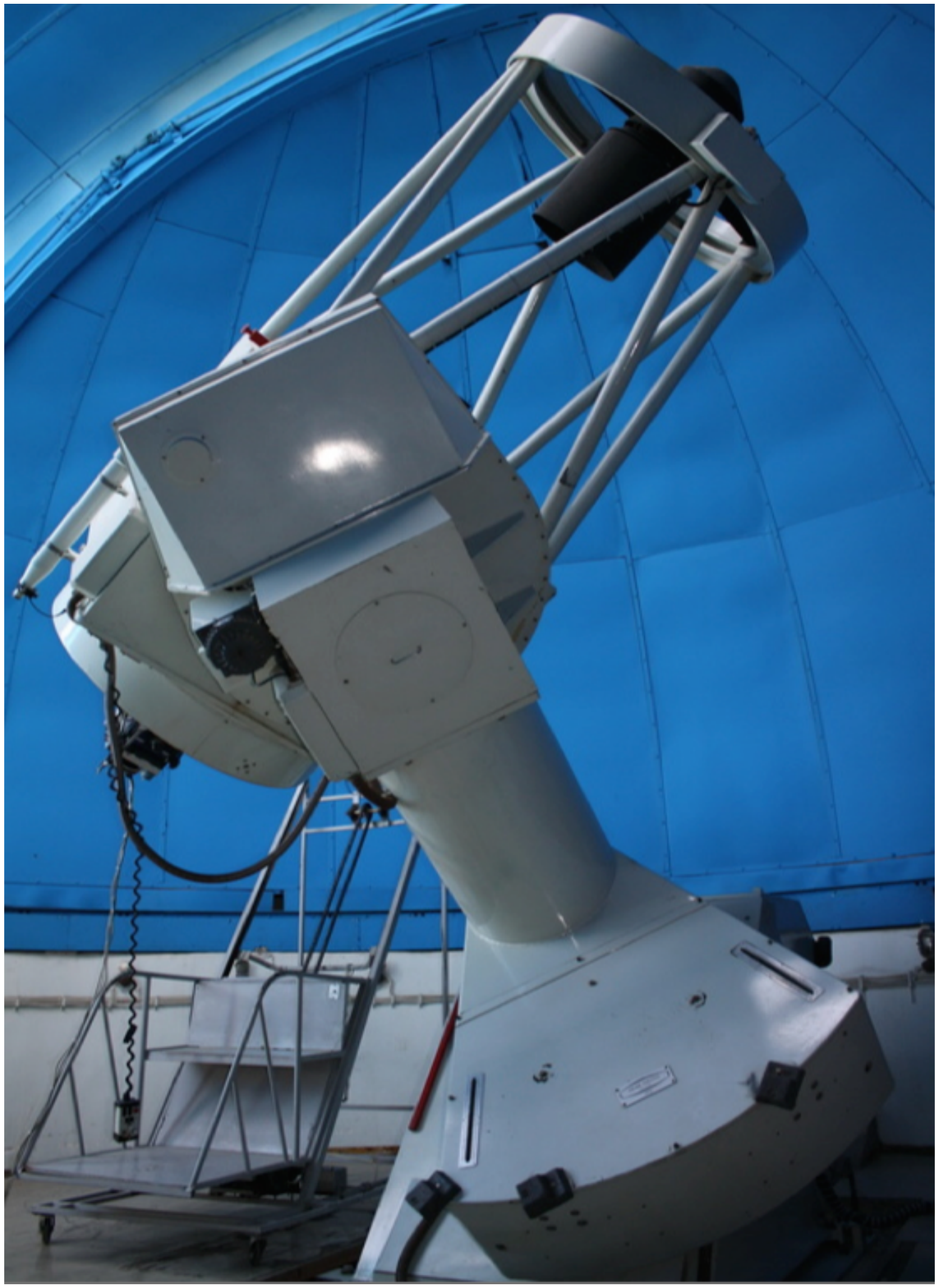} 
 \caption{The 1.2 m Cassegrain reflector telescope at Kryoneri
   Observatory, which is being retrofit and upgraded to detect lunar
   impacts in the framework of the NELIOTA project.}
   \label{fig1}
\end{center}
\end{figure}

NELIOTA aims to increase the number of detected faint lunar impacts, and
therefore increase the statistics to obtain their size distribution,
speeds, frequency, and characterize the impact ejecta. Using the 1.2m
Cassegrain reflector telescope at Kryoneri Observatory, manufactured and
installed in 1975 by the British company Grubb Parsons Co., Newcastle
(Figure~\ref{fig1}), we aim to push the detection limit for the first
time to $V= 12$ mag. Note, that the surface brightness of the earthshine
ranges between 12 m$_V$ arcsec$^{-2}$ (New Moon) and 17 m$_V$
arcsec$^{-2}$ (near Full Moon), with variations on the timescale of
hours of the order of 0.25 m$_V$ arcsec$^{-2}$ due to terrestrial
meteorology (Montanes-Rodriguez et al. 2007). Given the expected power
law size distribution of NEOs, we anticipate providing significant
numbers of small NEOs by detecting faint flashes. These data would be
valuable for characterizing the meteor environment and providing
guidelines to spacecraft manufacturers for protection of their vehicles,
as well as for future space mission planning.

The objective of NELIOTA is to design, develop and implement a highly
automated lunar monitoring system using existing facilities at the
National Observatory of Athens, Greece. For the first phase of the
project, DFM Engineering, Inc. will be retrofiting and upgrading the
electronics and mechanical parts of the 1.2m Kryoneri telescope, located
in the Northern Peloponnese, in Greece. A dual imaging instrument, also
designed and manufactured by DFM Engineering, Inc., along with two Andor
Zyla 5.5 sCMOS fast-frame cameras recording at 30 fps, will be used to
simultaneously monitor the non-illuminated lunar surface for impact
flashes and to reject cosmic rays. Our setup will provide a
field-of-view $\sim$17'x14'. Specialised software is being developed to
control the telescope and cameras, as well as to process the resulting
images to detect the impacts automatically. The NELIOTA system will then
publish the data on the web so it can be made available to the
scientific community and the general public. Following a 2 month
commissioning phase, there will be a 22 month observing campaign for NEO
impact flashes on the Moon. The impact events will be verified,
characterised and recorded. The 1.2m Kryoneri telescope will be capable
of detecting flashes far fainter than telescopes currently monitoring
the Moon.

\clearpage

\end{document}